\documentclass[floatfix,11pt,amssymb,amsfont,amsmath,superscriptaddress,float,longbibliography,nobalancelastpage]{revtex4-2}

\usepackage{here}
\usepackage{graphicx}
\usepackage{lmodern}
\usepackage{animate}
\usepackage{amssymb,amsfonts,amsmath}
\usepackage{braket}
\usepackage{xcolor}
\usepackage{dsfont}
\usepackage{adjustbox}
\usepackage{footnote}
\usepackage[colorlinks=true, linkcolor=red, breaklinks=true, citecolor=blue, urlcolor=red]{hyperref}

\usepackage{layouts}

\newcommand{\mc}{\mathcal}
\newcommand{\Rep}{\mathsf{Rep}}

\newcommand{\1}{\mathds{1}}

\newcommand{\update}[1]{#1}

\begin{document}

\preprint{APS/123-QED}

\title{Perfect Particle Transmission through Duality Defects}
\author{Atsushi Ueda}
\email{Atsuhi.Ueda@ugent.be}
\affiliation{Department of Physics and Astronomy, Ghent University, Krijgslaan 299, 9000 Gent, Belgium}
\author{Vic Vander Linden}
\affiliation{Department of Physics and Astronomy, Ghent University, Krijgslaan 299, 9000 Gent, Belgium}
\author{Boris De Vos}
\affiliation{Department of Physics and Astronomy, Ghent University, Krijgslaan 299, 9000 Gent, Belgium}
\author{Laurens Lootens}
\affiliation{Department of Applied Mathematics and Theoretical Physics, University of Cambridge,\\ Wilberforce Road, Cambridge, CB3 0WA, United Kingdom}
\author{Jutho Haegeman}
\affiliation{Department of Physics and Astronomy, Ghent University, Krijgslaan 299, 9000 Gent, Belgium}
\author{Paul Fendley}
\affiliation{Rudolf Peierls Centre for Theoretical Physics, Parks Rd, Oxford OX13PU,United Kingdom}
\affiliation{All Souls College, Oxford, OX14AL, United Kingdom}
\author{Frank Verstraete}
\affiliation{Department of Physics and Astronomy, Ghent University, Krijgslaan 299, 9000 Gent, Belgium}
\affiliation{Department of Applied Mathematics and Theoretical Physics, University of Cambridge,\\ Wilberforce Road, Cambridge, CB3 0WA, United Kingdom}

\date{\today}

\begin{abstract}
The theory of generalized symmetries has recently clarified how twisted sectors resolve the Callan-Rubakov paradox, where scattering of a charged particle by a magnetic monopole appeared to violate conservation laws. We study a more general setting of wavepackets that propagate across topological interfaces in quantum spin systems exhibiting non-invertible symmetries, and across duality defects coupling dual theories. In these scenarios, we find that the transmission is always perfect and a particle traversing the interface is converted into a nonlocal string-like excitation. We give a systematic way of constructing such a defect by identifying its Hilbert space with the virtual bond dimension of the matrix product operator representing defect lines. Our work provides a precise characterization of topological interfaces in perfect transmission phenomena and yields a lattice analogue of the solution to the monopole paradox in quantum field theory.
\end{abstract}

\maketitle

\section{Introduction.}
Quantum many-body systems often exhibit entirely counterintuitive behavior. One of such examples in quantum field theory is the well-known ``monopole paradox" originally proposed by Callan~\cite{Callan:1982ah,Rubakov:1982fp}. This paradox arises from a thought experiment: what happens if we throw a charged Weyl fermion at a magnetic monopole? Early studies found that the scattered particle is nothing like a Weyl fermion, possessing different and sometimes fractional quantum numbers. 
Naively, this suggests that the electron breaks into fractions under the influence of a magnetic monopole. As fractionalized electrons have not been observed in nature, a long debate about the interpretation of the results ensued.

\begin{figure}[tb]
    \centering

    \begin{minipage}[t]{0.25\textwidth}
        \centering
        \textbf{(a)}\\
        \includegraphics[width=40mm,page=1]{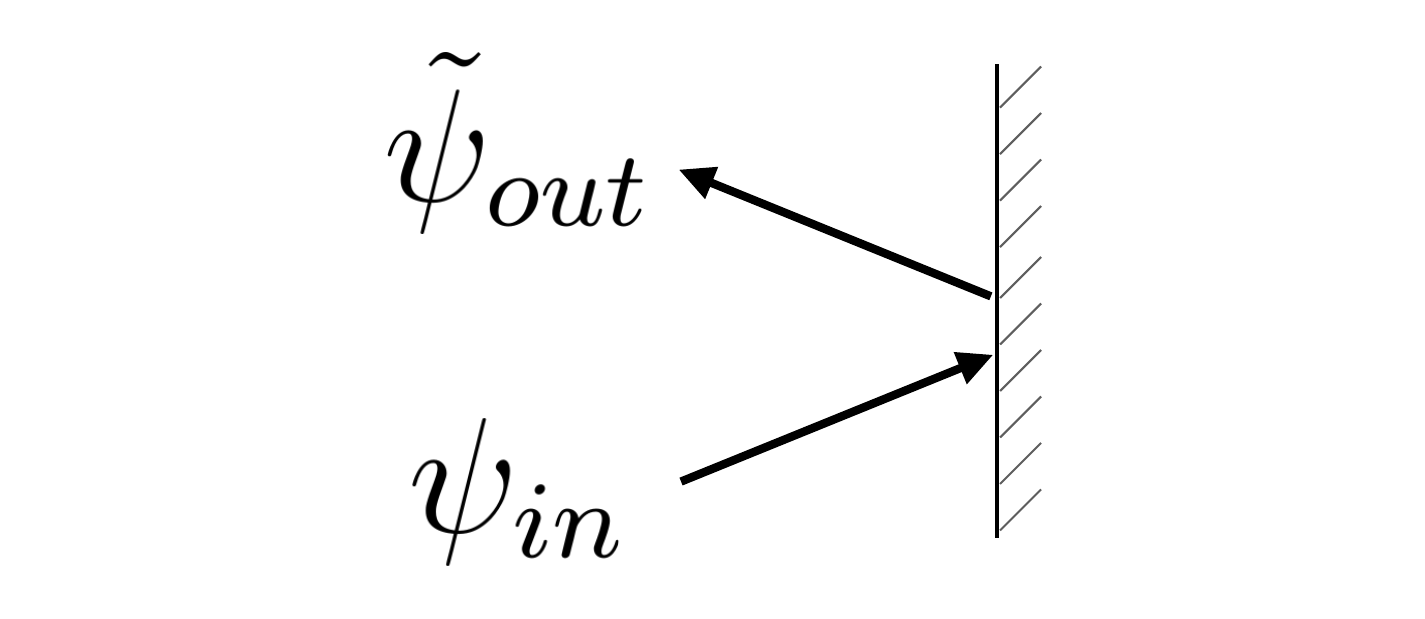}
    \end{minipage}
    \begin{minipage}[t]{0.25\textwidth}
        \centering
        \textbf{(b)}\\
        \includegraphics[width=40mm,page=2]{Fig1.pdf}
    \end{minipage}

    \vspace{2mm}

    \begin{minipage}[t]{0.25\textwidth}
        \centering
        \textbf{(c)}\\
        \includegraphics[width=40mm,page=3]{Fig1.pdf}
    \end{minipage}
    \begin{minipage}[t]{0.25\textwidth}
        \centering
        \textbf{(d)}\\
        \includegraphics[width=40mm,page=4]{Fig1.pdf}
    \end{minipage}

    \vspace{2mm}

    \begin{minipage}[t]{0.25\textwidth}
        \centering
        \textbf{(e)}\\
        \includegraphics[width=44mm,page=5]{Fig1.pdf}
    \end{minipage}

    \caption{\textbf{Monopole scattering and an impurity scattering model.}
    (a) Chiral fermion scattered by a Dirac monopole.
    (b) Fermion rotor model.
    (c) A schematic illustration of the system. Two systems with different Hamiltonians interact through the impurity, denoted by a black diamond.
    (d) We create a right-moving particle/wavepacket in the left medium on top of the many-body ground state.
    (e) When $H_{\text{L}}$ and $H_{\text{R}}$ are related by duality and separated by a topological impurity, the excitation propagates with perfect transmittance. The outgoing particle on the right appears to be a different particle, but can be described by the same particle with a topological string attached to the impurity.}
    \label{fig:illustration}
\end{figure}

Recent studies have addressed this question~\cite{vanBeest:monopole,Loladze:2025jsq,PhysRevLett.134.051602,vanBeest:2023mbs,Maldacena:1995pq,Affleck:1993np}. They revealed that the magnetic monopole serves as a conformal boundary between incoming and outgoing states shown in Fig.~\ref{fig:illustration}(a). This boundary directs the incoming fermion to a twisted sector by attaching a topological string to the scattered particle, which generates the fractional quantum number. This perspective is further supported by the fermion-rotor model~\cite{Polchinski:fermion_rotor}, which interprets the monopole as a quantum rotor impurity that scatters the fermion as shown in Fig.~\ref{fig:illustration}(b). In this case, the Weyl fermion is scattered perfectly by the impurity, but acquires a string and so behaves as a nonlocal particle. In this case, the \emph{unitarity puzzle}, a phenomenon
in which the S-matrix seemingly violates unitarity, was resolved by expanding the Hilbert space with particles that are connected to the impurity by a string. 

While the monopole paradox is a nice example that bridges quantum many-body dynamics and generalized symmetries in quantum field theory, it raises the question of whether this is a specific case. In this paper, we demonstrate that this phenomenon is ubiquitous in strongly interacting models. We show how to construct more generic and direct examples on a lattice. In these quantum spin chains, incoming particles are perfectly transmitted by the impurity and behave as if the original particle has disappeared from the spectrum. After illustrating this phenomenon with a simple example, we provide an intuitive picture of these seemingly exotic dynamics using the tensor-network framework. From the technical point of view, the results follow from the fact that both categorical symmetries and duality operators can be represented in the form of matrix product operator (MPO) algebras \cite{bultinck2017anyons,lootens2021matrix,lootens2023dualities,lootens2024dualities,Aasen:2016dop,aasen2020topologicaldefectslatticedualities}, and that those MPOs are sequential unitary quantum circuits \cite{lootens2025low,y9cd-dz5k}. The impurity degree of freedom is then associated with the dangling virtual space where the MPO ends. This provides a general answer to the unitarity puzzle. 

Our work substantially generalizes
analogous results in two-dimensional conformal field theory (CFT). Duality defects in conformal field theory (CFT) have been analyzed in depth \cite{FUCHS1999419,PETKOVA2001157,FUCHS2002353,PhysRevLett.93.070601,FROHLICH2007354,Bachas:2007td}, and particles passing through them are fully transmitted \cite{Thomas_Quella_2007}, possibly leaving a string in their wake. Our results apply exactly on the lattice, and neither criticality~\cite{PhysRevB.94.115125} nor integrability \cite{Sinha:2023hum} is needed. It is solely a consequence of the presence of generalized symmetries \cite{Nussinov:2009zz,Gaiotto:2014kfa,McGreevy:2022oyu,Shao:2023gho} and/or dualities \cite{Cobanera:2011wn,lootens2023dualities}. The identification of the impurity Hilbert space with the virtual bond of the MPO can easily be understood from the quantum to classical Euclidean path integral point of view. As demonstrated in \cite{Strangecorr,lootens2021matrix,Vanhove_2022}, classical statistical mechanical lattice models can be represented by tensor networks in which the defects correspond to MPOs that can be moved freely without changing the partition function. The crossing of such a topological MPO defect with the transfer matrix leads to an extra degree of freedom given by the virtual dimension of the MPO.

\section{Simple example. }
A key ingredient of the monopole paradox is that the Hilbert spaces for incoming and outgoing states are qualitatively different. In one non-anomalous example with spacetime a half-plane \cite{vanBeest:2023mbs,vanBeest:monopole}, the model is built from right-moving Weyl fermions with charges 3 and 4, and left-moving Weyl fermions with charges 5 and 0. Thus when a charge-3 particle is scattered from the monopole at the edge, having an outgoing charge-3 fermion seems to be {\textit{prohibited}}.

Similar situations arise in condensed-matter physics. For simplification, we consider two quantum spin chains described by Hamiltonians $H_{\text{L}}$ and $H_{\text{R}}$, coupled through an impurity $h_{\text{im}}$. The full Hamiltonian $H$ is given by
\begin{equation}
 H = H_{\text{L}} + H_{\text{R}} + h_{\text{im}}\ . \label{eq:Hamiltonian}
\end{equation}
This model is schematically shown in Fig.~\ref{fig:illustration} (c). 
As with the fermion-rotor model, we study the scattering problem by creating a wavepacket of a quasiparticle on top of the many-body ground state. We utilize the matrix-product-state framework \cite{PhysRevB.85.100408,10.21468/SciPostPhysLectNotes.7} to create a Gaussian wavepacket \cite{PhysRevResearch.3.013078,PRXQuantum.3.020316,PhysRevD.111.014504} of quasi-particle excitations, and then evolve it using the time-dependent variational principle \cite{PhysRevB.94.165116}. Generally speaking, $H_{\text{L}}$ and $H_{\text{R}}$ can be completely different Hamiltonians, and there is no guarantee that the particle can pass through the impurity. In fact, the original excitation on the left may correspond to a high-energy excitation on the right, making it energetically prohibited. However, when the spectrum of the two quantum spin systems matches, as is the case for dual systems,  an appropriately chosen junction allows the particle to pass completely through the impurity by changing its form. In the following, we see that the scattered particle is indeed the original wavepacket dressed with a topological line as illustrated in Fig.~\ref{fig:illustration}(d-e).

To elaborate, we take $H_{\text{L}}$ and $H_{\text{R}}$ to be two transverse-field Ising chains with ferromagnetic $XX$ and $ZZ$ couplings respectively: 
\begin{equation}
\begin{gathered}
    H_{\text{L}} = -\sum_{j=1}^{I-2} X_jX_{j+1} - g_{\text{L}}\sum_{j=1}^{I-1} Z_j\;,\qquad
    H_{\text{R}} = -\sum_{j=I+1}^{L} X_j - g_{\text{R}}\sum_{j=I+1}^{L} Z_jZ_{j+1},\\
    h_{\text{im}} = -X_{I-1}X_I -g_{\text{R}}\, Z_I Z_{I+1}, 
\end{gathered}
\label{HIsing}
\end{equation}
where $X_j$ and $Z_j$ act with Pauli matrices acting on site $j$ and trivially elsewhere. The impurity at site $I$ couples to the left and right chains by ferromagnetic $XX$ and $ZZ$ couplings respectively. Henceforth, we set $g_{\text{L}} = 4$ to ensure that the ground state on the left is polarized in the $Z$-direction.

This Hamiltonian \eqref{eq:Hamiltonian} built from \eqref{HIsing} has $\mathbb D_4 = \mathbb Z_4 \rtimes Z_2$ symmetry generated by $\prod_{j=1}^I Z_j$ and $\prod_{j=I}^{L+1} X_j$, which anticommute at the defect. Put differently, the defect degree of freedom transforms projectively under the $\mathbb Z_2 \times \mathbb Z_2$ symmetry, signaling a 't Hooft anomaly.


\begin{figure*}[tb]
    \centering

    \begin{minipage}[t]{0.49\textwidth}
        \centering
        \textbf{(a)}\\
        \includegraphics[width=88mm,page=1]{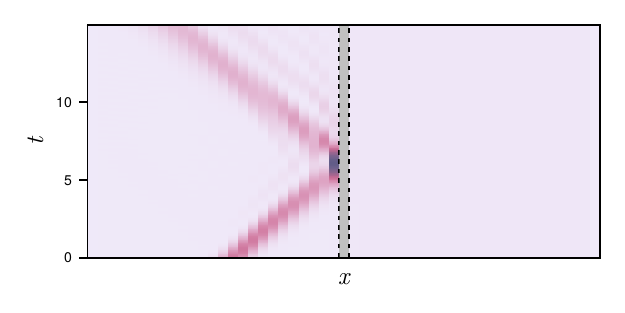}
    \end{minipage}
    \hfill
    \begin{minipage}[t]{0.49\textwidth}
        \centering
        \textbf{(b)}\\
        \includegraphics[width=88mm,page=2]{Fig2.pdf}
    \end{minipage}

    \vspace{2mm}

    \begin{minipage}[t]{0.49\textwidth}
        \centering
        \textbf{(c)}\\
        \includegraphics[width=88mm,page=3]{Fig2.pdf}
    \end{minipage}
    \hfill
    \begin{minipage}[t]{0.49\textwidth}
        \centering
        \textbf{(d)}\\
        \includegraphics[width=88mm,page=4]{Fig2.pdf}
    \end{minipage}

    \caption{\textbf{Numerical results.}
    The time evolution of the local magnetization $\langle Z_x\rangle$ for
    (a) $(g_L,g_R)=(4.0,2.0)$ and
    (b) $(g_L,g_R)=(4.0,4.0)$ with $(L,x_0,k) = (50,15,0.7\pi)$.
    The impurity site is represented by a gray shaded strip.
    (c) Transmittance rate of the wavepacket.
    (d) Perfect transmission of a domain wall of a ferromagnet to a Haldane chain.
    The video versions ``Supplementary\_Video\_1.gif'' are available in Supplementary Material and at
    \url{https://github.com/dartsushi/Video_scattering}.}
    \label{fig_TFIsing}
\end{figure*}

We next introduce a wavepacket of a $Z$ spin-flip excitation on the left chain, defined as \cite{PhysRevResearch.3.013078}
\begin{equation}\label{eq:wavepacket}
\widehat{W} = \sum_{j \in H_{\text{L}}} e^{-\frac{(j-j_0)^2}{2\sigma^2}} e^{ikj} X_j,    
\end{equation}
where $j_0$, $\sigma$ and $k$ set the packet’s center, width and mean momentum. The initial state is constructed by applying this operator to the nontrivial ground state of $H$:
$$|\psi(t=0)\rangle = \widehat{W} |\psi_{GS}\rangle.$$
The system evolves according to the original Hamiltonian in Eq.~\eqref{eq:Hamiltonian}, which drives the packet toward the interface. When the right chain is tuned to $g_{\text{R}} > g_{\text{c}} = 1$, it resides in the $Z$-ordered (ferromagnetic) phase, while the left chain with $g_{\text{L}} = 4$ is deep in the field-polarized (disordered) phase. Consequently, the interface connects two distinct gapped phases. Naively, one would expect a propagating spin-flip excitation from the left to be reflected, as a single-spin-flip excitation represents a high-energy excitation in the symmetry-breaking phase of $H_{\text{R}}$.

This is indeed the case for $g_{\text{R}}$\,=\,2 as shown in Fig.~\ref{fig_TFIsing}(a). The wavepacket, illustrated with pink, perfectly reflects at the interface and bounces back to the left edge. On the other hand, the outcome is different when $g_{\text{R}}$\,=\,$g_{\text{L}}$\,=\,4 as shown in Fig.~\ref{fig_TFIsing}(b): The wavepacket passes through the impurity with perfect transmittance. This behavior occurs only when $g_{\text{L}}$\,=\,$g_{\text{R}}$, as shown in Fig.~\ref{fig_TFIsing}(c). For $g_{\text{L}}=4$, the
scattering is purely reflective for $g_{\text{R}}<2$ and $g_{\text{R}}>6$. Indeed, the single-particle dispersion relation is $\epsilon_k = 2\sqrt{1+g^2-2\cos(k)}$, so that there is no overlap of energy spectrum when $|g_{\text{L}}-g_{\text{R}}|>2$.

In the perfectly transmitting case $g_{\text{L}}=g_{\text{R}}$, the state changes character after scattering. It is no longer a spin-flip excitation but a domain wall: it flips the spins as it moves along in the symmetry-breaking phase, as apparent from Fig.~\ref{fig_TFIsing}(b). Such a domain wall is indeed a low-energy state in this phase.  As we will see, this domain-wall excitation can be understood as the spin-flip excitation dressed with a topological string $\Pi_{i=1}^{x(t)} X^R_i$ with $x(t)$ being the center of the wavepacket at time $t$. This behavior is exactly that described in Fig.~\ref{fig:illustration}, raising the question: What is special about this point $g_{\text{L}}=g_{\text{R}}$?

\section{Perfect transmission and topological defects.} 

A short answer to this question is the following: perfect transmission occurs when two mutually dual Hamiltonians $H_{\text{L}}$ and $H_{\text{R}}$ are separated by a topological ``duality" defect. In the Ising example above, $H_{\text{L}}$ and $H_{\text{R}}$ are related by the Kramers--Wannier (KW) transformation when $g_{\text{L}}$\,=\,$g_{\text{R}}$~\cite{KW_1941,Fradkin:1978th}, and the corresponding duality defect has long been known \cite{Schutz1993,Oshikawa:1996dj}. In an effective field-theory description, a local spin excitation turns into a disorder excitation when scattering through a duality defect. Since the theories are dual, the two excitations must have the same gap and dispersion relation. 

The unitary duality operator $U_{KW}$ provides a convenient way for understanding the perfect transmission. This operator is nicely expressed in terms of a unitary quantum circuit/MPO \cite{Aasen:2016dop,lootens2025low} 
\begin{align}
 \includegraphics[width=0.5\linewidth,page=3]{fig_KWoperator.pdf},
 \label{Udef}
\end{align}
where $\oplus$ and the black dot represent the XOR gate and the Greenberger-Horne-Zeilinger (GHZ) state~\cite{GHZ_1989}, respectively. While the KW duality is often implemented via a non-invertible operator, in \eqref{Udef} we have lifted the virtual degree of freedom to a physical one. Acting with $U_{KW}$ does not alter the energy spectrum in this enlarged Hilbert space. More importantly, its application to the subsystem also does not change the physics. 

This operator allows us to demonstrate perfect transmission in the impurity system by unitarily transforming it to a uniform system. The action of $U_{KW}$ on the bulk parts of the Hamiltonian is simple to understand in terms of the local operators
\begin{align}
h_{j}=-X_j X_{j+1} - g Z_{j+1}\;,\qquad h'_j = -X_{j}- g Z_j Z_{j+1} \ .
\label{hdef}
\end{align}
so that at the perfectly transmitting point $g_{\text{L}}=g_{\text{R}}\equiv g$ and
\begin{align}
H_{\text{L}} = -gZ_1+\sum_{j=1}^{I-2} h_j\ ,\qquad\quad
H_{\text{R}} = \sum_{j=I+1}^{L-1} h'_j\ .
\end{align}
It is then straightforward to check that conjugation by $U_{KW}$ relates these terms as
\begin{equation}
h'_{j+1}=U_{KW}^\dagger h_j U_{KW}\qquad\iff\qquad    \begin{gathered}
\includegraphics[width=0.4\linewidth,page=1]{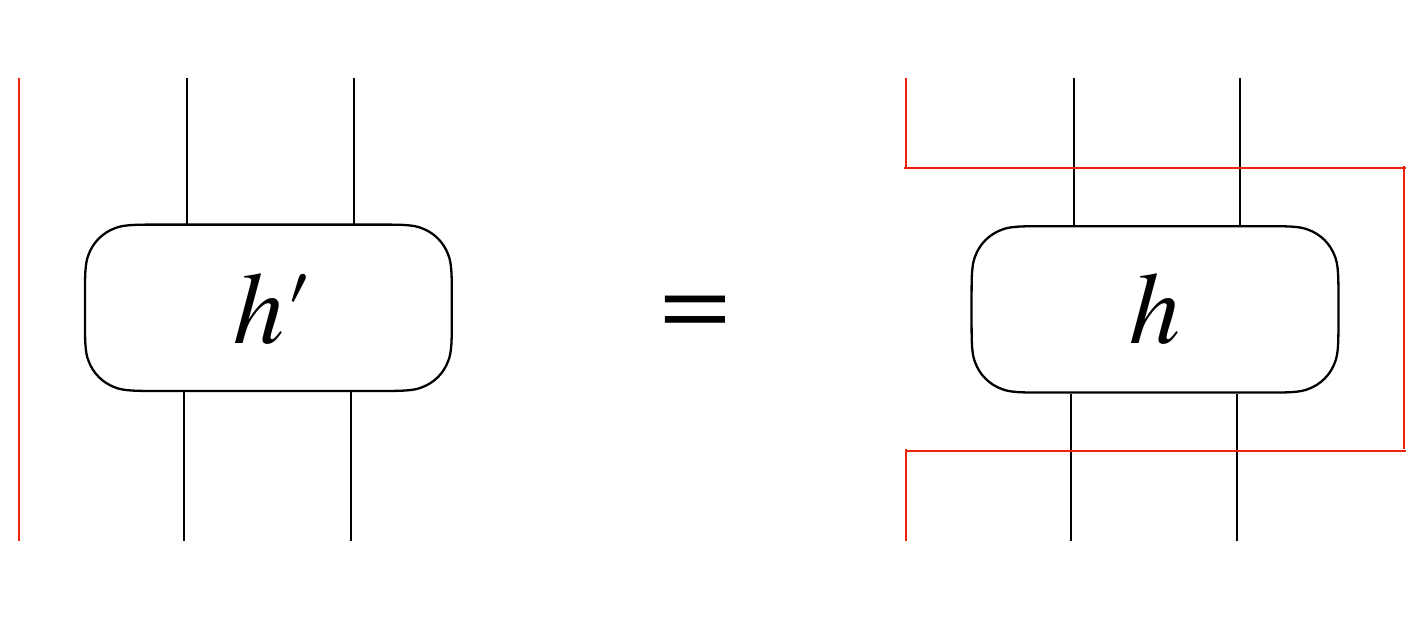}
\end{gathered}
\label{UhU}
\end{equation}
The reason for the shift in $j$ is that an extra degree of freedom was introduced to make the duality transform unitary.
The impurity is defined by defining $U_{KW}$ to act only on the right half of the chain, so that the extra site required is now at site $I$. The impurity Hamiltonian is then
\begin{align}
\begin{gathered}
    \includegraphics[width=0.4\linewidth,page=2]{eq_h.pdf}
\end{gathered}
\label{himpdef}
\end{align}
This diagram gives a completely general and constructive way of defining topological defects for any generalized symmetry and/or duality represented by an MPO. Note that this also works for systems with long-range interactions; in that case, a string between the interacting parts will emerge for the impurity Hamiltonian. 

\begin{figure}[tb]
    \centering
   \includegraphics[width=180 mm]{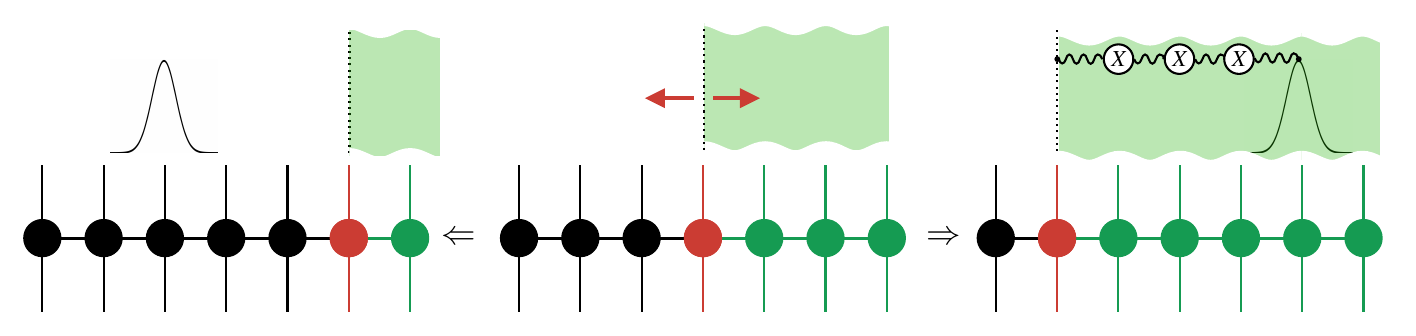}
    \caption{\textbf{Equivalence between perfect transmittance and topological defects.} Moving the topological impurity is a unitary transformation and does not alter the Hamiltonian spectrum. A localized spin-flip wavepacket transforms into a domain wall.}
    \label{fig:defect_moving}
\end{figure}

We have thus shown that the Hamiltonian~\eqref{eq:Hamiltonian} with perfect transmission can be unitarily transformed to
\begin{align}
    {H}  = U^\dagger_{KW}(\tilde{H}\otimes \mathds 1_{L+1})U_{KW},\label{eq:half_unitary}
\end{align}
where $\tilde{H}=-\sum_{j=1}^{L-1} h_j$ is a uniform Hamiltonian of length $L$. This transformation can be schematically represented as
\begin{equation*}
    \includegraphics[width=0.8\linewidth]{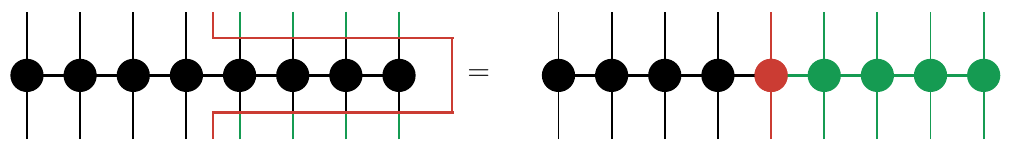}.
\end{equation*}
The impurity site, marked by the red line, originates from the extra state introduced by $U_{KW}$. 
This invariance allows the topological impurity to move freely from right to left as shown in Fig.~\ref{fig:defect_moving}. This movement is similar to drawing a curtain. When the wavepacket propagates to the right, it behaves as if no impurity were present, and passing through the impurity is analogous to hiding behind the green curtain. The topological $X$ string from the interface is just that of the duality MPO or, in more traditional language, the disorder operator. The insertion of the identity implies that no energy will get stuck on the interface when the wavepacket alters its degrees of freedom. 

\update{From this perspective, we can observe the equivalence between perfect transmission and duality. If \textit{any} wavepacket can pass perfectly through the impurity in both directions, we conjecture that there exists a duality transformation represented by a specific MPO that leaves the spectrum invariant (up to degeneracies). The reason is simple: perfect transmission for arbitrary excitations is only possible when the two theories have a one-to-one spectral correspondence. In this context, the impurity degrees of freedom ensure that unitarity is preserved and the degeneracies match. More generally, the perfect transmission of particles in a particular subsector only signals a partial duality between $H_{\text{L}}$ and $H_{\text{R}}$. Note that an accidental coincidence of energy gaps, as observed in resonant tunneling, is not sufficient, since a general wavepacket overlaps with a very large number of eigenstates. We expect this conjecture to extend to higher dimensions, as we discuss in the conclusion and outlook. From an experimental perspective, it is interesting to observe string breaking by measuring the impurity site.}

As the transformation \eqref{eq:half_unitary} is unitary, the spectra of $H$ and $\tilde{H}\otimes \mathds 1_{L+1}$ must be identical. The latter has an obvious doubled degeneracy for all energy levels arising from the spin at $L+1$, and a corresponding $SU(2)$ symmetry. The full symmetry of $H$ is therefore $\mathbb Z_2 \times SU(2)$.


\section{A non-integrable example}

The transformation Eq.~\eqref{eq:half_unitary} is not restricted to free-fermion models such as Ising, or even integrable models. We apply it here to the spin-1 Heisenberg chain, which realizes the gapped Haldane phase~\cite{Haldane:1982rj,Haldane_1983}. While this phase is now known as a canonical example of symmetry-protected topological (SPT) order~\cite{PhysRevB.80.155131,Chen:2012ctz,Senthil:2014ooa}, it is dual to the usual ferromagnet through the Kennedy--Tasaki (KT) transformation~\cite{KT_2,KT_1992}. This transformation can be represented with a bond dimension 2 MPO with local tensor $T_{ij\alpha\beta}=\delta_{ij}\tilde{\sigma}^i_{\alpha\beta}$ with $\tilde{\sigma}^i=(\sigma^x,i\sigma^y,\sigma^z)$, where the first and the last two indices refer to the physical and virtual spaces, respectively~\cite{lootens2023dualities}.  we then {\em define} 
the dual and impurity Hamiltonians by replacing $U_{KW}$ in \eqref{UhU} and \eqref{himpdef} with the analogous unitary operator here. Defining $h_j= S_j\cdot S_{j+1}$, we obtain
\begin{equation}
\begin{gathered}
h'_j =-S^x_jS^x_{j+1}+e^{i\pi S^x_j}S^y_jS^y_{j+1}e^{i\pi S^z_{j+1}}-S^z_jS^z_{j+1}\;,\cr
h_{\mathrm{im}} = S^x_{I-1}\sigma^z_{I} S^x_{I+1} + iS^y_{I-1}\sigma^y_{I}S^y_{I+1}e^{i\pi S^z_{I+1}} - S^z_{I-1}\sigma^x_{I}Z_{I+1}
\end{gathered}
\end{equation}
It is important to note that the impurity is a qubit distinct from the spin-1 physical space. This qubit is similar to the edge mode of the AKLT ground state, which transforms projectively under the $\mathbb{Z}_2\times\mathbb{Z}_{2}$ symmetry that is being gauged by the KT transform~\cite{KT_2,KT_1992,AKLT_1987,Pollmann_2012}. 
We then define the full Hamiltonian as \eqref{eq:Hamiltonian} as before, where here $H_{\text{L}}=\sum_{j=1}^{I-1} h_j$ and $H_{\text{R}}=\sum_{j=I+1}^L h'_j$\;. 

By construction, this $H$ can be unitarily transformed into a uniform Hamiltonian. Transmission across the impurity must therefore be perfect. 
The scattering process is illustrated in Fig.~\ref{fig_TFIsing}~(d), where the left and right sides have been interchanged for clarity. \update{The elementary excitation in the ferromagnetic phase is a domain wall. A possible wavepacket realising such a domain wall can be created by replacing $X_j$ in Eq.~\eqref{eq:wavepacket} by $\mu_j=\prod_{k<j} S_k^x$.} The domain wall, created in the ferromagnetic phase, propagates through the impurity and becomes invisible. This invisible wave then flips the edge spins of the SPT phase when it reaches the right boundary. The wavepacket becomes invisible in $S^z$ in the SPT phase because it becomes a non-local string operator $\tilde{S}^z_n = e^{i\pi \sum_{j=1}^{n} S^{z}_j}S_n^{z}$~\cite{Nijs_1989}. This string attachment is inferred from correlation functions, where $\langle S^{z}_iS^{z}_j\rangle = 0,$ when $i<I<j$ but $\langle \tilde{S}^{z}_iS^{z}_j\rangle \neq 0$. This is analogous to what was observed in the fermion-rotor model. A similar effect occurs for the spin-spin correlation function of the Ising model as reported in Ref.~\cite{Oshikawa:1996dj,Brehm:2015lja}.

\section{Topological defects from symmetric matrix product operators.}


\begin{figure}[tb]
    \centering
        \textbf{(a)}\\
        \includegraphics[width=88 mm, page=1]{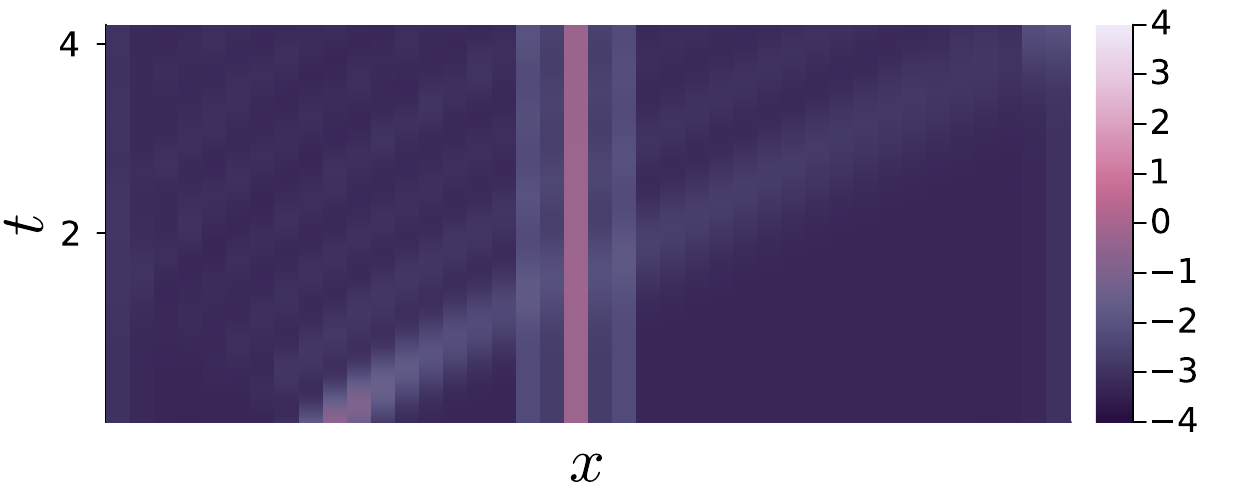}\\
            \textbf{(b)}\\
        \includegraphics[width=88mm, page=2]{Fig4.pdf}\\
    \caption{\textbf{$\Rep(\mc S_3)$ model. }The time evolution for the $\Rep(\mc S_3)$ model with $(g_{\text{L}},g_{\text{R}})=(3.0,3.0)$ of (a) the local energy and (b) the local magnetization $\braket{Z_x}$. Total transmittance is visible, as well as the spin flip on the impurity. The topological string is invisible due to the symmetry-respecting ground state.}
    \label{fig:Rep(S3)}
\end{figure}
At criticality, the Ising model is self-dual, and the KW operator is promoted to a symmetry of the Hamiltonian~\footnote{An extra on-site Hadamard gate is needed if we use $U_{KW}$ defined in the previous section}. In this scenario, $H_{\text{L}}$ and $H_{\text{R}}$ are identical, resulting in a uniform system with a non-invertible impurity at the center. This concept can be generalized to any system with MPO symmetry. 

Here, we consider the following $\Rep(\mc S_3)$ symmetric model \cite{lootens2024dualities,10.21468/SciPostPhys.16.5.127,Bhardwaj:2024wlr}, which indeed corresponds to the case $H_{\text{L}}=H_{\text{R}}$ through a non-trivial \update{MPO symmetry}:
\begin{align}
    H_{\text{L}} &= \sum_{i=2}^{I-2} (\sigma^x_{i-1} (\mathds 1 + \sigma^z)_{i} \sigma^x_{i+1}) - g_{\text{L}}\sum_{i = 1}^{I-1} \sigma^z_i,\nonumber\\
    H_{\text{R}} &= \sum_{i=I+2}^{L-1} (\sigma^x_{i-1} (\mathds 1 + \sigma^z)_{i} \sigma^x_{i+1}) - g_{\text{R}}\sum_{i = I+1}^{L} \sigma^z_i,\nonumber\\
    h_{\text{im}} &= \frac{1}{2} \sigma^x_{I-2} (\mathds 1+\sigma^z)_{I-1} (\sigma^x + \sqrt{3}\sigma^y)_I \sigma^x_{I+1} \nonumber\\
    &+ \frac{1}{2} \sigma^x_{I-1} (\sigma^x - \sqrt{3}\sigma^y)_I (\mathds 1+\sigma^z)_{I+1} \sigma^x_{I+2}.\nonumber
\end{align}
The MPO is constructed from the local tensor $T_{ij\alpha\beta} = \delta_{ij} \rho(i)_{\alpha\beta}$, with $\rho(1) = \sigma^x, \rho(2) = (\mathds 1 + \sqrt{3} \sigma^z)/2$, \update{and squares to the sum of itself, the identity operator and a global on-site $\sigma^z$}. Pulling it through half the chain as in (\ref{himpdef}) yields the topological boundary represented by $h_{\text{im}}$. Perfect particle transmission of a wavepacket (same as Eq.~\eqref{eq:wavepacket}) is illustrated in Fig \ref{fig:Rep(S3)}. Moreover, as the particle passes the boundary, it can be observed that the impurity spin flips. This is a consequence of invisible string attachment, as the topological string for this model is absorbed by the symmetric ground state. \update{It is also worth stressing that a single-spin flip itself is not an elementary excitation but rather a composite particle, but this does not affect the property of perfect transmission.}

\section{Conclusion and Outlook.}
In this paper, we demonstrated how perfect transmission arises from the presence of duality defects. By applying a unitary MPO that represents the duality transformation or symmetry to a subsystem of the Hamiltonian, we derive a lattice impurity model with perfect transmission. The impurity degrees of freedom emerge from the virtual legs of the MPO, revealing hidden insights into the duality or symmetry of the MPO. From the point of view of generalized symmetries, this MPO construction makes the connection between topological defects, 't Hooft anomalies and degeneracies in the spectrum very explicit. These impurities are topological and attach a topological string to the scattered particle. Conversely, we conjecture that any interface between two theories that allows for perfect transmission implies the existence of a duality between the theories at both sides.

 \update{This viewpoint opens up a promising route to uncovering non-trivial duality through quantum dynamics in numerical simulations and experiments. Even without {\it{a priori}} knowledge of duality, observing perfect transmission at a special point in the parameter space, just as in Fig.~\ref{fig_TFIsing} (c), can be taken as a smoking gun evidence of the existence of either a duality or a generalized symmetry.} Recently, such dualities and generalized symmetries have been exploited to construct more efficient DMRG algorithms for simulating quantum spin chains \cite{lootens2025entanglement}; within that framework, Hamiltonians with topological defects can be simulated using translationally invariant ones without impurities, therefore yielding a direct path to reproducing all results in this paper.
\begin{figure}[bt]
    \centering
    \includegraphics[width=88 mm]{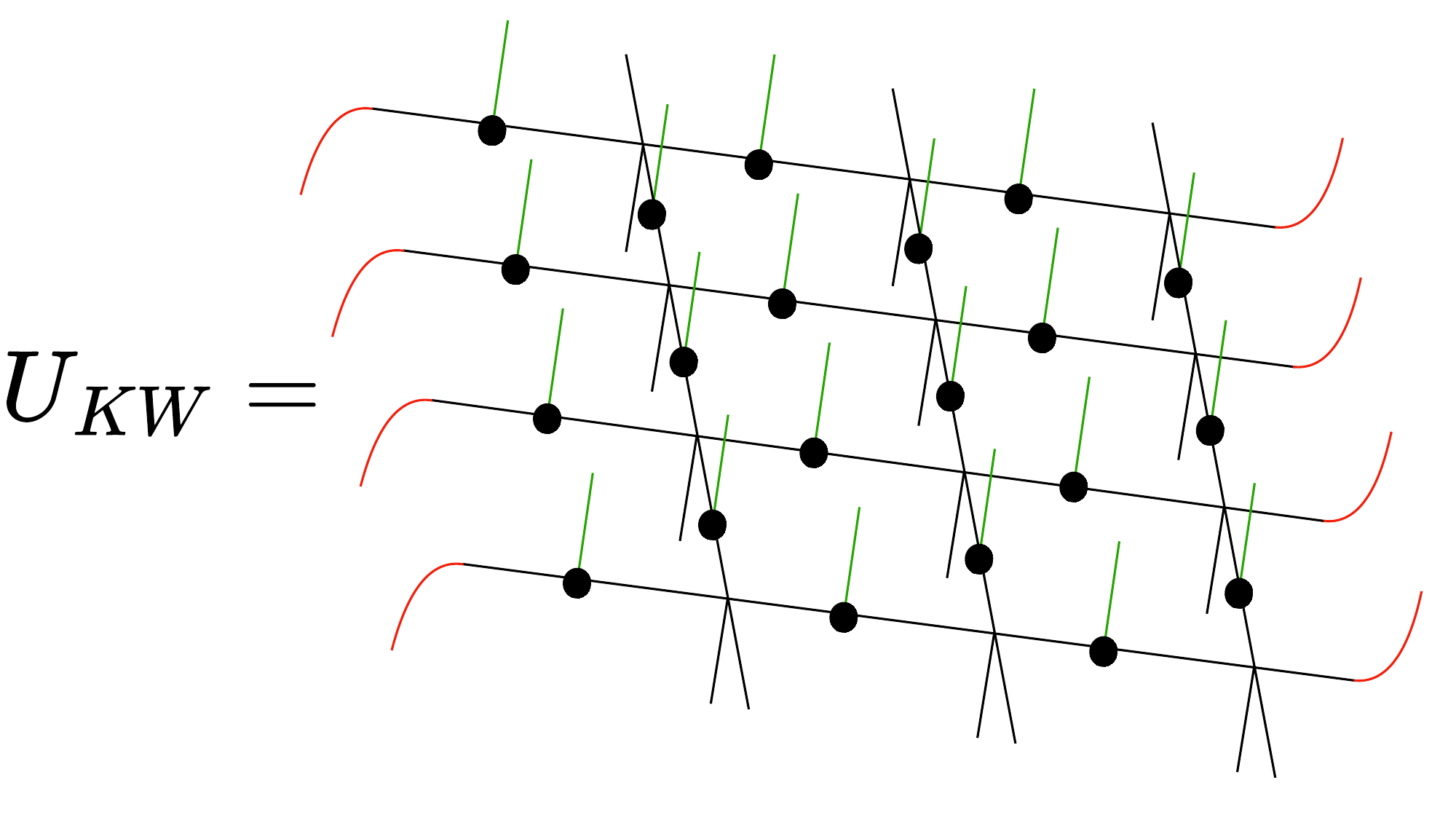}
    \caption{\textbf{The Kramers--Wannier duality in two dimensions.} This maps the transverse-field Ising model on a square lattice to the model on links. The virtual bonds denoted with red lines become a duality defect line.}
    \label{fig:2d_duality}
\end{figure}
Our model serves as a simple lattice example of the physics underlying the monopole paradox. In relativistic fermionic models with $\mathrm{U(1)}$ symmetry, a symmetric MPO, if it exists, should exhibit $ \mathrm{U(1)} $ irreducible representation virtual degrees of freedom at low energies. Indeed, the impurity manifests as a bosonic rotor within the fermion-rotor model. Looking ahead, we conjecture a duality between charge 3 and 4 Weyl fermions and charge 5 and 0 Weyl fermions. However, studying chiral fermions on a lattice remains challenging. We aim to explore this using recent advancements in tensor-network-based chiral fermion formalisms~\cite{Haegeman:2024qgf,PhysRevResearch.6.043059,PhysRevLett.133.116501} in the near future.

It is worth noting that our recipe applies to any dimension, not just one. In two dimensions, one can construct a topological line defect by applying a projected entangled-pair operator (PEPO) to a subsystem as shown in Fig.~\ref{fig:2d_duality}. Such a PEPO represents the intertwiner between 2-D dual theories when following the gauging procedure promulgated in \cite{PhysRevX.5.011024}. For the case of the transverse-field Ising model, gauging the $\mathbb{Z}_2$ symmetry leads to the toric code with an external field $H_{\text{R}} = -\sum_{i\in \mathrm{link} }Z_i -g\sum_{v\in\mathrm{vertex}}{A_v}$, where $A_v = \Pi_{i\in v} X_i$ with the product over the links that touch the vertex $v$. 
A single spin flip of the symmetric Ising phase that crosses the defect interface is then converted into a charge excitation, carrying a string that extends to the interface.
Due to the non-locality of this string, it cannot be detected by single-site operators. Nevertheless, one can detect its existence by a non-local Wilson-loop operator such as a product of $X$ operators surrounding it. The numerical simulation of this scattering process is available in \url{https://github.com/dartsushi/Video_scattering}.

\section{Acknowledgment}
We thank the Isaac Newton Institute for Mathematical Sciences, Cambridge, for support and hospitality during the programme ``Quantum field theory with boundaries, impurities, and defects", where work on this paper was undertaken after an inspiring talk by David Tong (workshop supported by EPSRC grant EP/Z000580/1). We also thank Victor Vanthilt and Lander Burgelman for helpful discussions. Finally, we thank the referees for the insightful comments that significantly improved our manuscript. 

\section{Author Contributions Statement}
All authors contributed to the conceptual framework, to the tensor network computations, and to the writing of the manuscript.

\section{Funding Statements}

A.U. is supported by FWO Junior Postdoctoral Fellowship (grant No. 3E0.2025.0049.01) and Watanabe Foundation.
V.L, B.V, and F.V acknowledge funding from EOS (Grant No. 40007526), IBOF (Grant No. IBOF23/064), and BOF-GOA (Grant No. BOF23/GOA/021), the  UKRI grant EP/Z003342/1 and the European Research Council (ERC) under the European Union’s Horizon 2020 program (Grant Agreement No. 101125822). L.L. is supported by an EPSRC Postdoctoral Fellowship (grant No. EP/Y020456/1). P.F.\ by
EPSRC Grant no. EP/X030881/1.
\section{Competing Interests Statement}
The authors declare no competing interests.


\section{Figure Legends/Captions}
\underline{\textbf{Figure 1}}

 \textbf{Monopole scattering and an impurity scattering model.}(a) Chiral fermion scattered by a Dirac monopole. (b) Fermion rotor model. (c) A schematic illustration of the system. Two systems with different Hamiltonians interact through the impurity, denoted by a black diamond. (d) We create a right-moving particle/wavepacket in the left medium on top of the many-body ground state. (e) When $H_{\text{L}}$ and $H_{\text{R}}$ are related by duality and separated by a topological impurity, the excitation propagates with perfect transmittance. The outgoing particle on the right appears to be a different particle, but can be described by the same particle with a topological string attached to the impurity. 

\underline{\textbf{Figure 2}}

\textbf{Numerical results.} The time evolution of the local magnetization $\langle Z_x\rangle$ for (a) $(g_L,g_R)=(4.0,2.0)$ and (b) $(g_L,g_R)=(4.0,4.0)$ with $(L,x_0,k) = (50, 15,0.7\pi)$. The impurity site is represented by a gray shaded strip. (c) Transmittance rate of the wavepacket.  (d) Perfect transmission of a domain wall of a ferromagnet to a Haldane chain.
    The video versions “Supplementary\_Video\_1.gif" are available in Supplementary Material and \url{https://github.com/dartsushi/Video_scattering}.

\underline{\textbf{Figure 3}}

\textbf{Equivalence between perfect transmittance and topological defects.} Moving the topological impurity is a unitary transformation and does not alter the Hamiltonian spectrum. A localized spin-flip wavepacket transforms into a domain wall.

\underline{\textbf{Figure 4}}

\textbf{$\Rep(\mc S_3)$ model. }The time evolution for the $\Rep(\mc S_3)$ model with $(g_{\text{L}},g_{\text{R}})=(3.0,3.0)$ of (a) the local energy and (b) the local magnetization $\braket{Z_x}$. Total transmittance is visible, as well as the spin flip on the impurity. The topological string is invisible due to the symmetry-respecting ground state.

\underline{\textbf{Figure 5}}

\textbf{The Kramers--Wannier duality in two dimensions.} This maps the transverse-field Ising model on a square lattice to the model on links. The virtual bonds denoted with red lines become a duality defect line.

\providecommand{\noopsort}[1]{}\providecommand{\singleletter}[1]{#1}%

\section{Methods}

In physics, a duality is the statement that two apparently different theories are, in fact, equivalent descriptions of the same physical theory. A canonical example is the Kramers--Wannier duality of the Ising model \cite{KW_1941}. This duality relates the spontaneously symmetry-broken (SSB) phase to a corresponding point in the symmetric phase. The statement can be made precise: in the classical Ising model, the partition functions at dual temperatures $T$ and $T^*$ are identical. Similarly, for the quantum transverse-field Ising chain, the Hamiltonians $H_{\text{L}}$ and $H_{\text{R}}$ introduced in the main text have exactly the same spectrum.

In this sense, a duality is not a change of the physical theory itself, but a change in how the theory is represented \cite{Cobanera:2011wn}. The same theory may be written as an SSB Hamiltonian, or equivalently as a $\mathbb{Z}_2$-symmetric Hamiltonian. The operators in the two descriptions can nevertheless look very different. In particular, the symmetry of the two descriptions are in general not invariant under the duality; the action of operators which leave the theory invariant will also look different. As discussed in the main text, a domain-wall excitation in the SSB phase appears, at first sight, to be quite distinct from a single spin-flip excitation in the symmetric phase. However, the two are related by a unitary transformation. Thus, once the theory is fixed, its ground states and excitations are fixed as well. The duality only provides a different language, or a different set of labels, for the same physical objects. Depending on the physical theory, this duality may provide an easier description in calculating certain physical quantities such as correlation functions, or in determining properties such as phase transitions or integrability.

An interesting situation arises when one tries to represent the same theory using two different descriptions at once, for instance a symmetric description on the left half and an SSB description on the right half. There is then a mismatch at the interface, and one needs a patch that makes the two descriptions consistent. At first sight, this patch may appear arbitrary. However, once the state and the symmetry are fixed, the mathematics fixes the patch as well.

We start off from a physical theory with a certain symmetry described by a fusion category $\mc{C}$, and choose an initial description of the theory by specifying an invertible bimodule category $\mc{R}$ over it. With this, a particular realization of the theory can be written down by making use of the dual fusion category $\mc{D} = \mc{C}^*_\mc{R}$. We interpret $\mc{R}$ as a particular representation of the physical theory; a $\mc{C}$-symmetric Hamiltonian can be written down with $(\mc{D}, \mc{R})$. A duality then boils down to changing this representation $\mc{R}$ to a different one $\mc{R}'$, while keeping $\mc{D}$ fixed \cite{lootens2023dualities}. This causes the symmetry to change from $\mc{C}$ to $\mc{C}' = \mc{D}^*_{\mc{R}'}$. This is consistent with our earlier statement that the realization of the symmetry changes under dualities. The patch that lives between two theories represented by $\mc{R}$ and $\mc{R}'$ is determined by the module functors of the category
\begin{equation*}
    \mathrm{Fun}_{\mc{D}}(\mc{R},\mc{R}').
\end{equation*}

With this, we have the following natural interpretation of the patch. We first start off from one representation of the theory $\mc{R}$ for the whole system. We then change the representation to $\mc{R}'$ on, say, the right half of the system. We view this as the action of the duality on the right half of the system, and the interface here is the left end of the duality. Crossing this interface results in the duality action, and is thus the patch that sticks dual theories together in a consistent manner. In modern literature, this is referred to as generalized (half-)gauging \cite{Shao:2023gho}.

For instance, the $\mathbb{Z}_2$ symmetry of the Ising model is encoded by starting from
\begin{equation*}
    (\mc{C}, \mc{R}, \mc{D}) = (\mathsf{Vec}_{\mathbb{Z}_2}, \mathsf{Vec}, \mathsf{Rep}(\mathbb{Z}_2)).
\end{equation*}
The two representatives of the Ising model are described by
\begin{equation*}
    \mathcal{R} = \mathsf{Vec}
    \qquad \text{and} \qquad
    \mathcal{R}' = \mathsf{Rep}(\mathbb{Z}_2).
\end{equation*}
For a fixed coupling constant $g$, one of them will represent a theory in the SSB phase, and the other in the symmetric phase. The categorical framework then determines how to connect these two representations through
\begin{equation*}
    \mathsf{Fun}_{\mathcal{D}}(\mathsf{Vec},\mathsf{Rep}(\mathbb{Z}_2)),
\end{equation*}
from which in the Ising model we understand the Kramers--Wannier duality to be defined by.

The explicit tensor-network representation of these functors is discussed in Ref.~\cite{lootens2023dualities, lootens2024dualities,lootens2025low}. Here, the transformation between $\mathcal{R}$ and $\mathcal{R}'$ is constructed from the building-block objects called $F$-symbols, in this case specifically of the category $\mathsf{Fun}_{\mc{D}}(\mc{R},\mc{R}')$. The intuition is that an $F$-symbol is the elementary local move which changes the order in which topological lines are fused. When the duality defect is dragged across the lattice, this local recoupling move is applied repeatedly. The MPO implementing the duality is therefore obtained by placing these local $F$-moves along the chain. In this way, the consistency conditions required from the categorical data are automatically satisfied on tensor networks, rather than imposed by hand \cite{bultinck2017anyons, lootens2021matrix}. In the Kramers--Wannier case, the MPO tensor for the unitary transformation $U_{\mathrm{KW}}$ is determined by the $F$-symbols associated with $\mathsf{Fun}_{\mathcal{D}}(\mathsf{Rep}(\mathbb{Z}_2),\mathsf{Vec})$. This is given by
\begin{equation*}
    W =
    \sum_{a,b\in\{0,1\}}
    |a\rangle\,|a+b\rangle\langle b|\,\langle a+b| ,
\end{equation*}
where $a+b$ is understood modulo $2$. The four legs of this tensor act on the left, bottom, top, and right bonds, respectively. In particular, the first and last legs are virtual indices, and are simply qubits. It is straightforward to check that this is the same MPO as the one used in the main text.

The categorical data determine not only the duality map, but also how local Hamiltonian terms transform when one chooses the representation $\mathcal{R}$. This is expressed by the pulling-through equation
\begin{equation*}
    \includegraphics[width=\linewidth]{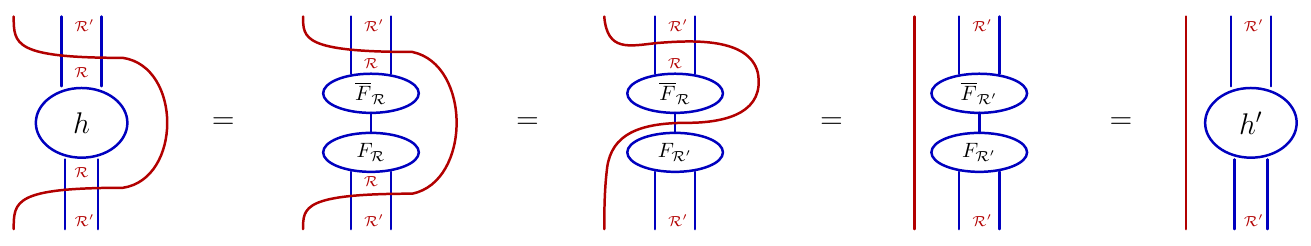}
\end{equation*}
which says that the patch between $\mathcal{R}$ and $\mathcal{R}'$ can be moved freely through the tensor network, changing the representation along the way. In this sense, it behaves as a topological line.

Importantly, as shown in Ref.~\cite{lootens2025low}, the duality transformation can always be represented in a unitary form. This is the mathematical origin of the unitary transformation introduced in the main text. The same construction applies beyond the Ising example. For the Haldane chain, one fixes
\begin{equation*}
    (\mathcal{D},\mathcal{R},\mathcal{R}')
    = (\mathsf{Rep}(\mathbb{Z}_2\times\mathbb{Z}_2),\mathsf{Vec},\mathsf{Rep}^{\psi}(\mathbb{Z}_2\times\mathbb{Z}_2)),
\end{equation*}
to represent the SPT phase as its Kennedy--Tasaki dual, while for the $\mathsf{Rep}(\mc{S}_3)$ model one takes
\begin{equation*}
    (\mathcal{D},\mathcal{R},\mathcal{R}')
    =
    (\mathsf{Rep}(\mc{S}_3), \mathsf{Rep}(\mathbb{Z}_2),\mathsf{Rep}(\mathbb{Z}_2)) .
\end{equation*}
In the latter case, the representation is unchanged, $\mathcal{R}=\mathcal{R}'$, so the corresponding unitary acts as a symmetry of the Hamiltonian. From this point of view, a symmetry can be understood as a trivial duality: This duality does not even alter the representation. Nevertheless, the topological interface itself can be non-trivial, as shown in the main text. 

The framework above has a natural physical interpretation. The ability to pull the duality map through the tensor network means that one is free to choose the most convenient representation locally. The fact that the map can be chosen unitary means that the underlying physical content is independent of this choice. In this sense, duality maps do not change the theory itself; they relabel the same physical degrees of freedom, and this relabeling is done through categorical data.

Finally, we note that the observation that in the case of Kramers-Wannier duality, the impurity degree of freedom transforms projectively can be understood in this categorical language as well. Kramers-Wannier is obtained by the gauging of a global $\mathbb Z_2$ symmetry. In general, gauging a $\mathsf{Vec}_G$ symmetry leads to a $\mathsf{Rep}(G)$ symmetry, with corresponding bimodule category $\mathsf{Vec}$. When thought of as a module category over $\mathsf{Vec}_G \boxtimes \mathsf{Rep}(G)^{\text{op}}$ instead, it labels an SPT phase, as noted in \cite{Fechisin:2023odt}.

\section{Data Availability}
The data used in the figures are available in \url{https://github.com/dartsushi/Video_scattering}.
\section{Code Availability}
The codes used in the figures are available in \url{https://github.com/dartsushi/Video_scattering}.


\end{document}